\documentclass[epj]{svjour}
\usepackage{graphics}
\newcommand{\bm}[1]{\mbox{\boldmath $#1$}}
\newcommand{\bms}[1]{\mbox{\scriptsize\boldmath $#1$}}
\begin{document}
\title{Geometric study for the Legendre duality of generalized entropies 
and its application to the porous medium equation}
\author{Atsumi Ohara
}                     
\offprints{}          
\institute{Department of Systems Science, Osaka University, 1-3 Machikane-yama, Toyonaka, Osaka, 560-8531 JAPAN \\ \email{ohara@sys.es.osaka-u.ac.jp}}
\date{Received: / Revised version: }
%
\abstract{
We geometrically study the Legendre duality relation that plays an 
important role in statistical physics with the standard or 
generalized entropies.
For this purpose, we introduce dualistic structure defined by 
information geometry, and discuss 
concepts arising in generalized thermostatistics, 
such as relative entropies, escort distributions and modified expectations.
Further, a possible generalization of these concepts in a certain direction 
is also considered.
Finally, as an application of such a geometric viewpoint, 
we briefly demonstrate 
several new results on a behavior of the solution to 
the nonlinear diffusion equation called the {\em porous medium equation}.
} 
\authorrunning{A. Ohara}
\titlerunning{Geometric study for the Legendre duality of 
generalized entropies and its application.}
\maketitle

\section{Introduction}

In recent decades study of physical or artificial systems 
not obeying the usual Boltzmann-Gibbs statistical mechanics 
has received an increasing attention. 
For examples of such systems see 
\cite{BS,Tsallis04,Frank,Chavanis} and the references therein.
Their common nature would be that the Boltzmann distribution 
does not correspond to their equilibriums.
One of main research directions to overcome 
the difficulty of analysis for such systems
is generalizing the notion of entropies within the 
framework of statistical physics. 

In this generalization, the Legendre duality relation is still of 
fundamental importance and is required to prescribe 
a link between intensive and extensive parameters.
In statistics this nice structure has been well exploited via 
{\em information geometry} \cite{Amari,AN} 
for the standard exponential family, 
and the results are successfully applied mainly to statistical estimation, 
information theory, learning theory and so on.

The purpose of this paper is to study the Legendre duality relation 
of generalized entropies from 
information geometric viewpoints and provide 
new insights and tools with this field by showing their 
usefulness through several applications.

In the aspect of geometric structure with Legendre duality, 
there may be at least two major directions to generalize the notion of 
entropies from the standard one.
In section 2 we characterize the difference of these two methods 
in terms of a pair of representing functions for distributions.
The one method always fixes one of representing functions 
to the identity map while the other varies both.
The former leads to the geometry induced by the {\em Bregman divergence}.
The latter includes the {\em $\alpha$-geometry} as a special case.

From section \ref{shortrev} to \ref{GrFl}, we discuss the 
Tsallis statistics, its generalization and applications.
Section \ref{shortrev} presents a short review of the relation 
of the $\alpha$-geometry with the Tsallis statistics and emphasizes 
its importance.
In section \ref{escprob} we reconsider the construction of 
the $\alpha$-geometry by the {\em affine surface theory} \cite{Kurose1,Kurose2,NS} 
as preparation to a more generalized setup.
The geometrical relation of two manifolds of the ordinary and the escort 
distributions are discussed.
Section \ref{GCE} proposes a generalization of the 
$\alpha$-geometric structure and the 
associated divergences using a certain class of convex functions.
It is seen that the centro-affine immersion \cite{NS} is essential to conserve 
the dualistic structure.
In section \ref{app1} and \ref{GrFl}, we demonstrate applications of the 
introduced geometric notions to exploit the properties of 
generalized entropies.
Section \ref{app1} shows the relation between modified averages 
(expectations) and 
convexities, which plays an important role in minimizing relative entropies 
under the average constraints.
In section \ref{GrFl}, we prove that the trajectory of the gradient flow for 
the $\alpha$-divergence is a geodesic curve and it possesses 
several constants of motion.

Finally in sections \ref{GEU} and \ref{NDE}, we introduce the so-called 
Bregman divergence \cite{Bregman}, 
the associated generalized entropy and geometry behind these quantities 
studied in \cite{Eguchi,MTKE,Naudts02,Naudts041}.
The feature is that the linear averages of the extensive physical quantities 
naturally appear in this setup.
As an application, 
we show several new results on the behavior of the solutions to 
the {\em porous medium equation} (PME). 
The behavior is characterized in terms of geometric concepts induced 
on a generalized exponential family called the {\em $q$-Gaussian densities}.
This family can be proved an invariant manifold for the PME.
Further, we show that the trajectory of the solution on the manifold 
coincides with a geodesic curve with respect to the one of the
dual affine connections.
In addition the convergence rate to the manifold is evaluated.
For this part, a full description with complete proofs can be found in 
\cite{OW08}.

\section{Statistical model and dualistic structure}
\label{sec2}
Let $p_\zeta(x)=p(x;\zeta)$ be a probability distribution 
for random variable $x$ 
(or, density function for the continuous random variable)
parametrized by a finite-dimensional parameter vector 
$\zeta=(\zeta^1,\cdots,\zeta^n) \in {\cal Z}$, 
where ${\cal Z}$ is a certain domain in ${\bf R}^n$.
We call the set of $p_\zeta$  
{\em the statistical model} and denote it by ${\cal M}$.
The concrete examples in this paper are 
the probability simplex (\ref{simpmodel}) 
and the $q$-Gaussian densities (\ref{qGaussmodel}).
We usually assume that ${\cal M}$ satisfies several 
{\em regularity conditions} such as smoothness of the map 
$\zeta \mapsto p_\zeta$, commutativity of 
integrations and differentiations and so on. 
See \cite{Amari,AN} for details.

It is known that the standard Boltzmann-Gibbs- \break Shannon (BGS) entropy 
is maximized on the the Boltzmann distribution 
(exponential family) with the expectation 
constraint of the Hamiltonian.
Similarly, each generalized entropy is maximized 
on the corresponding statistical model with the constraint.
See, for example, Remark \ref{rem_proj} in section \ref{GEU}.
This is the major reason that motivates us to study structure of specific 
statistical models focusing on the Legendre duality 
of generalized entropies.

Information geometry is a convenient framework for this purpose.
In order to introduce the geometric structure on 
the statistical model, we define the following quantities:
\[
	g_{ij}(\zeta):=\int \frac{\partial L(p_\zeta)}{\partial \zeta^i}
		\frac{\partial L^*(p_\zeta)}{\partial \zeta^j} dx,
\]
\[
 \Gamma_{ij,k} (\zeta) 
	:= \int \frac{\partial^2 L(p_\zeta)}{\partial \zeta^i 
		\partial \zeta^j} 
		\frac{\partial  L^*(p_\zeta)}{\partial \zeta^k} dx,
\]
\[
 \Gamma^*_{ij,k} (\zeta) 
	:= \int \frac{\partial L(p_\zeta)}{\partial \zeta^k}
		\frac{\partial^2 L^*( p_\zeta)}
		{\partial \zeta^i \partial \zeta^j} dx.
\]
Here, a pair of one-to-one and smooth functions $L(u)$ and $L^*(u)$ 
on $u \ge 0$ are called 
{\em representing functions} for the distribution $p_\zeta$, which 
determines the Legendre duality such as pairs of dual coordinate systems 
(physically, extensive and intensive parameters) or potential functions 
conjugate each other.
We use $g:=(g_{ij})$ as a Riemannian metric on ${\cal M}$ and 
\[
	\Gamma^{k}_{ij}:= \sum_{l=1}^n g^{kl}\Gamma_{ij,l}, \quad 
	\Gamma^{*k}_{ij}:= \sum_{l=1}^n g^{kl}\Gamma^*_{ij,l}
\]
as the components for two affine connections $\nabla$ and $\nabla^*$ 
on ${\cal M}$, 
where $g^{ij}$ is the component of the inverse matrix of $g$.
Then the above definitions imply that the following duality relation of 
the connections \cite{Amari,AN} holds:
\begin{equation}
	\partial_i g_{jk}=\Gamma_{ij,k}+\Gamma^{*}_{ik,j},
\label{drel}
\end{equation}
which is equivalent with (\ref{dualrel}) in Appendix A.
This relation is important for the geometric study of the Legendre duality.

The typical cases are classified as follows:
\begin{description}
\item[i)] The standard information geometry corresponding to the BGS entropy 
and Kullback-Leibler relative entropy is derived by
\[
	L(u)= u, \quad L^*(u)=\ln u.
\]
\item[ii)] The {\em $\alpha$-geometry} corresponding to 
the Havrda-Charvat-Tsallis entropy 
(\ref{HCTentropy}) and Tsallis relative entropy (\ref{Tre}) utilizes 
\[
	L(u)=L^{(\alpha)}(u):=\frac{2}{1-\alpha}u^{(1-\alpha)/2}, 
	\quad L^*(u)=L^{(-\alpha)}(u).
\]
This class, its generalization and applications are discussed from 
section \ref{shortrev} to \ref{GrFl}.
\item[iii)] Information geometry called the {\em U-geometry} \cite{Eguchi} 
is corresponding to the {\em Bregman-type divergences} 
\cite{Eguchi,MTKE,Naudts02,Naudts041,Naudts043} 
and the associated generalized entropies. 
It is reproduced from   
\[
	L(u)= u, \quad L^*(u)= \ln_\phi (u),
\]
where $\ln_\phi$ is a generalized logarithmic function defined by, e.g., 
(\ref{glog}) \cite{Naudts02,Naudts041,Naudts043}.
This class and its applications are discussed in section 
\ref{GEU} and \ref{NDE}.
\end{description}
Because $L(u)=u$ in the cases i) and iii), the linear average naturally 
appears and plays an important role.
In these cases, the corresponding connections $\nabla$ and $\nabla^*$ are 
called (generalized) {\em mixture} and {\em exponential connections}, 
respectively.
On the other hand, in the cases i) and ii) the obtained Riemannian metric $g$ 
coincides with the Fisher information, i.e.,
\begin{equation}
	g_{ij}(\zeta)=\int p_\zeta 
	\frac{\partial \ln p_\zeta}{\partial \zeta^i}
	\frac{\partial \ln p_\zeta}{\partial \zeta^j} dx.
\label{cndF}
\end{equation}
This is a very important point in applying the geometry 
to statistical inference.
We see from (\ref{cndF}) that the relation
\[
	\frac{dL}{du}\frac{dL^*}{du}=\frac{1}{u}
\]
should be satisfied for $g$ to be the Fisher information matrix.
Hence, the Riemannian metric in the case iii) is not generally 
the Fisher metric. 
However, it is physically interpreted as a susceptance matrix 
via the linear average (See section \ref{GEU}).

\section{Review of Tsallis entropy via alpha-geometry}
\label{shortrev}

Let $\mathcal{S}^n$ denote the $n$-dimensional probability simplex, i.e., 
\begin{equation}
	\mathcal{S}^n := \left\{ \bm{p}=(p_i) \left| \;
		p_i > 0, \; \sum_{i=1}^{n+1} p_i =1 
	\right. \right\}
\label{simpmodel}
\end{equation}
and $p_i, i=1,\cdots,n+1 $ denote
probabilities of $n+1$ states.
The set ${\cal S}^n$ is an example of the statistical model with 
parameters $p_i, i=1,\cdots,n$.
The function $S_q$ defined on $\bar{\mathcal{S}}^n$, 
the closure of $\mathcal{S}^n$, for a real parameter 
$q (\not=0 \mbox{ nor } 1)$ by
\begin{eqnarray}
 S_q(\bm{p}) &:=& -k \sum_{i=1}^{n+1} (p_i)^q \ln_q p_i \nonumber \\
&=&  \frac{k}{1-q} \left( \sum_{i=1}^{n+1} (p_i)^q - 1 \right)
\label{HCTentropy}
\end{eqnarray}
is called the Havrda-Charvat-Tsallis (HCT) entropy \cite{HC,Tsallis88}, 
where $\ln_q$ is the {\em $q$-logarithmic function} defined by 
$\ln_q x =(x^{1-q}-1)/(1-q)$.
Note that the HCT entropy converges to the BGS 
entropy when $q$ goes to one.
Hereafter, the positive constant $k$ is set to one for the sake of simplicity.
The HCT entropy is concave if $q>0$ and convex if $ q < 0$.
It does not satisfy the additivity, i.e., it holds that
\begin{equation}
 S_q(\bm{p} \otimes \bm{r})=S_q(\bm{p})+S_q(\bm{r})+(1-q)S_q(\bm{p})S_q(\bm{r}).
\label{nonextensivity}
\end{equation}
for $\bm{p}=(p_i) \in \bar{\mathcal{S}}^n$, 
$\bm{r}=(r_j) \in \bar{\mathcal{S}}^m$ and 
$\bm{p} \otimes \bm{r}:= (p_ir_j) \in \bar{\mathcal{S}}^{nm+n+m}$.
The relation (\ref{nonextensivity}) is called {\em nonextensivity}.
See, for details, a recent review paper \cite{Tsallis04}.

For several reasons the following quantity $K_q$ is introduced in 
\cite{RK,AD,BPT,Shiino98} as a relative entropy 
between two probability distributions 
$\bm{p}$ and $\bm{r}$ in $\mathcal{S}^n$, which is of the form:
\begin{eqnarray}
 K_q(\bm{p},\bm{r}) 
&:=& - \sum_{i=1}^{n+1} p_i \ln_q \left( \frac{r_i}{p_i} \right) \nonumber \\
&=&\frac{1}{1-q} \left( 1 - \sum_{i=1}^{n+1} (p_i)^q(r_i)^{1-q} \right).
\label{Tre}
\end{eqnarray}
When $q$ is positive, $K_q(\bm{p},\bm{r}) \ge 0$ and the equality holds 
if and only if $\bm{p}=\bm{r}$.
Note that $K_q$ converges to the Kullback-Leibler divergence 
as $q$ goes to one.
For the uniform distribution $\bm{u}=(u_i)$ with 
$u_i=1/(n+1)$ for all $i=1,\cdots n+1$, it holds
\[
 K_q(\bm{p},\bm{u})=\left( \frac{1}{n+1} \right)^{1-q} \{S_q(\bm{u})-S_q(\bm{p})\}.
\]
Hence, if $q>0$, the maximizing the HCT entropy 
$S_q(\bm{p})$ is equivalent to
minimizing $K_q(\bm{p},\bm{u})$.

On the other hand, the quantity called the 
{\em $\alpha$-divergence} \cite{Amari,AN} $D^{(\alpha)}$ has been used 
in mathematical statistics, which is defined with a real parameter 
$\alpha (\not= \pm 1)$ by
\begin{equation}
	D^{(\alpha)}(\bm{p},\bm{r}):=
\frac{4}{1-\alpha^2} \left\{ 1 - \sum_{i=1}^{n+1}
(p_i)^{(1-\alpha)/2}(r_i)^{(1+\alpha)/2} \right\}
\label{alphaD}
\end{equation}
for two probabilities $\bm{p}$ and $\bm{r}$ in $\mathcal{S}^n$. 
By equating 
\[
	q:=(1-\alpha)/2,
\]
we see that the Tsallis relative entropy $K_q$
coincides with the $\alpha$-divergence 
on $\mathcal{S}^n$ up to constant, i.e, 
\begin{equation}
 D^{(\alpha)}(\bm{p},\bm{r})
=\frac{1}{q} K_q(\bm{p},\bm{r}).
\end{equation}
Note that $D^{(\alpha)}$ is nonnegative regardless to $\alpha$ and 
positive if and only if $\bm{p} \not= \bm{r}$.
It also converges to the Kullback-Leibler divergence when 
$\alpha \rightarrow -1$ 
and it is convex with respect to $\bm{p}$ and $\bm{r}$ if $-1< \alpha <1$.

It is known that the $\alpha$-divergence induces a differential geometric 
structure on ${\cal S}^n$ with a Riemannian metric and 
an affine connection denoted by $g$ and $\nabla^{(\alpha)}$, respectively. 
We call it the {\em $\alpha$-geometry} \cite{Amari,AN}.
While the way to induce from the $\alpha$-divergence is omitted here, 
the resultant componentwise expressions of $g$ and $\nabla^{(\alpha)}$ are 
as follows:
Let $\partial_i$ be a natural basis tangent vector field on ${\cal S}^n$ defined by
\begin{equation}
	\partial_i:=\frac{\partial}{\partial p_i}
			-\frac{\partial}{\partial p_{n+1}}, \quad
			i=1,\cdots,n.
\label{basis}
\end{equation}
Then, the induced Riemannian metric is nothing but the Fisher information 
matrix, i.e.,
\begin{eqnarray}
	g_{ij}(\bm{p})&:=& g (\partial_i, \partial_j)
=\frac{1}{p_i}\delta_{ij}+ \frac{1}{p_{n+1}} \label{Rmetric} \\
&=&\sum_{k=1}^{n+1} p_k \frac{\partial \log p_k}{\partial p_i}
\frac{\partial \log p_k}{\partial p_j}, \quad i,j=1,\cdots,n.
\nonumber
\end{eqnarray}
The induced affine connection $\nabla^{(\alpha)}$ is called the 
{\em $\alpha$-connection}, which is represented in its coefficients by
\begin{equation}
	 \Gamma^{(\alpha)k}_{ij}(\bm{p}) 
=\frac{1+\alpha}{2}
\left(-\frac{1}{p_k}\delta_{ij}^k + p_k g_{ij}\right), 
\quad i,j,k=1,\cdots,n,
\label{alpha}
\end{equation}
where $\delta_{ij}^k$ is equal to one if $i=j=k$ and zero otherwise.
Then we have its covariant derivatives by
\[
	\nabla^{(\alpha)}_{\partial_i} \partial_j
		=\sum_{k=1}^{n} \Gamma^{(\alpha)k}_{ij} \partial_k.
\]

There are two specific features for the 
$\alpha$-geometry on ${\cal S}^n$ induced in such a way.
First, the triple $({\cal S}^n,g,\nabla^{(\alpha)})$ is a
{\em statistical manifold} (See Appendix for its definition), 
i.e., we can confirm that the following holds:
\begin{equation}
	Xg(Y,Z)=g(\nabla^{(\alpha)}_X Y,Z)+g(Y,\nabla^{(-\alpha)}_X Z)
\label{dc}
\end{equation}
for arbitrary vector fields $X, Y$ and $Z$ on ${\cal S}^n$.
Thus, $\nabla^{(\alpha)}$ and $\nabla^{(-\alpha)}$ are {\em mutually dual}.
The relation (\ref{dc}) is closely related with the Legendre duality.

Another feature is that $({\cal S}^n,g,\nabla^{(\alpha)})$ is a 
{\em manifold with constant curvature} $\kappa=(1-\alpha^2)/4=q(1-q)$, i.e., 
it holds that
\[
	 R^{(\alpha)}(X,Y)Z= \kappa \{g(Y,Z)X - g(X,Z)Y \},
\]
where $R^{(\alpha)}$ is the Riemann-Christoffel curvature with respect to 
$\nabla^{(\alpha)}$.
Because of this property the $\alpha$-divergence meets the 
{\em modified Pythagorean relation} 
for $\bm{p},\bm{q}$ and $\bm{r}$, which form a 
``right triangle" on ${\cal S}^n$ with respect to $g$ and 
$\nabla^{(\pm \alpha)}$, i.e.,
\begin{proposition}
\label{prop_mP}
 Let $\gamma^{(\alpha)}$ and $\gamma^{(-\alpha)}$ be respectively 
the $\nabla^{(\alpha)}$-geodesic joining $\bm{p}$ and $\bm{q}$, 
and the $\nabla^{(-\alpha)}$-geodesic joining $\bm{q}$ and $\bm{r}$.
If $\gamma^{(\alpha)}$ and $\gamma^{(-\alpha)}$ are orthogonal at $\bm{q}$ 
with respect to $g$, then it holds
\begin{eqnarray}
	D^{(\alpha)}(\bm{p},\bm{r}) &=& 
 D^{(\alpha)}(\bm{p},\bm{q}) + D^{(\alpha)}(\bm{q},\bm{r}) \nonumber \\
&&-\kappa D^{(\alpha)}(\bm{p},\bm{q})D^{(\alpha)}(\bm{q},\bm{r}).
\label{mP}
\end{eqnarray}
\end{proposition}

This relation is quite important in studying the properties of 
the HCT entropy $S_q$ because 
its nonextensivity (\ref{nonextensivity}) is straightforwardly 
derived from (\ref{mP}).
It means that nonflatness ($\kappa \not=0$) of the 
manifold ${\cal S}^n$ is geometrically interpreted as a direct 
cause of the nonextensivity \cite{Ohara07}.
Further, (\ref{mP}) ensures the uniqueness of the equilibrium distribution 
minimizing the Tsallis relative entropy $K_q$ with constraints given 
in terms of the normalized $q$-expectation \cite{Ohara07}.
(See also the discussion in section \ref{app1}.)

\section{Escort distribution from a viewpoint of affine immersion}
\label{escprob}
In the previous section the $\alpha$-geometry is introduced from the 
$\alpha$-divergence.
Another way to construct the $\alpha$-geometry on ${\cal S}^n$ is 
using the {\em affine immersion} \cite{NS} of ${\cal S}^n$ into 
${\bf R}^{n+1}$ equipped with the standard flat connection $D$.
The advantage of this method is that 
the escort probability naturally appears accompanying with the setup 
and its geometrical meaning is elucidated.
Hereafter, we assume that $1>q=(1-\alpha)/2>0$ to simplify the discussion.
Several concepts of affine immersion are summarized in the appendix.
For detail, see the references.

Let $\bm{\theta}=(\theta^i), i=1,\cdots,n+1$ be the standard 
coordinate system of the vector space ${\bf R}^{n+1}$ 
with respect to $\{o; e_1,\cdots,e_{n+1}\}$, 
the origin as zero vector and the natural basis vectors.
Denote by ${\bf R}^{n+1}_+$ the positive orthant of ${\bf R}^{n+1}$.

Consider the immersion $f$ of ${\cal S}^n$ into 
${\bf R}^{n+1}_+$ by
\begin{equation}
	f: \bm{p}=(p_i) \mapsto \bm{\theta}=(\theta^i)=(L^{(\alpha)}(p_i)), 
	\quad i=1,\cdots,n+1,
\label{immersion}
\end{equation}
where the representing function $L^{(\alpha)}$ is defined by
\begin{equation}
	L^{(\alpha)}(t):= \frac{2}{1-\alpha} t^{(1-\alpha)/2}=\frac{1}{q}t^q.
\label{theta}
\end{equation}
Note that $f({\cal S}^n)$ is  
a level surface $\psi(\bm{\theta})=2/(1+\alpha)$ 
in ${\bf R}^{n+1}_+$ of the function defined by
\begin{equation}
 \psi(\bm{\theta}) := \frac{2}{\alpha +1} \sum_{i=1}^{n+1}
 \left(\frac{1-\alpha}{2} \theta^i \right)^{2/(1-\alpha)}
=\frac{1}{1-q} \sum_{i=1}^{n+1}
 \left(q \theta^i \right)^{1/q}.
\label{potential}
\end{equation}
By the assumption for the range of $q=(1-\alpha)/2$, the function $\psi$ 
is convex with the positive definite Hessian matrix on ${\bf R}^{n+1}_+$.

At this stage, we still have a freedom of choosing the transversal vector $\xi$.
The freedom induces realizations of the different geometric structure of 
$({\cal S}^n,h,\nabla)$.
Here we take $\xi$ as
\begin{equation}
	\xi:=  \sum_{i=1}^{n+1} \xi^i \frac{\partial}{\partial \theta^i},
	\quad \xi^i= -(1-q)q \theta^i.	
\label{trans}
\end{equation}
This choice of $\xi$ is derived by 
\begin{equation}
	\xi=-\frac{1}{d\psi(E)}E
=-\frac{1}{\sum_{i=1}^{n+1} (\partial \psi/\partial \theta^i) E^i}E,
\label{scale}
\end{equation}
where $E=\sum_{i=1}^{n+1}E^i \partial/\partial \theta^i$ 
is the vector field defined to satisfy
\begin{equation}
	\frac{\partial^2 \psi}{\partial \theta^i \partial \theta^j} X^i E^j
	=\sum_{i=1}^{n+1} \frac{\partial \psi}{\partial \theta^i} X^i
	=d\psi(X),
\label{normal}
\end{equation}
for an arbitrary vector field 
$X=\sum_{i=1}^{n+1} X^i \partial/\partial \theta^i$ on ${\bf R}^{n+1}_+$.
Hence, if $X$ is tangent to $f({\cal M})$, then the right-hand side of 
(\ref{normal}) vanishes. 
However, since the Hessian 
$(\partial^2 \psi/\partial \theta^i \partial \theta^j)$ is 
positive definite, $E$ and $\xi$ are guaranteed transversal to $f({\cal M})$.
Further, we see 
from (\ref{trans}) that the immersion $(f,\xi)$ is {\em centro-affine} 
with a scaling constant $q(1-q)$.

As summarized in the appendix, the centro-affine immersion realizes 
a statistical manifold $({\cal S}^n,h,\nabla)$ with constant curvature.
Actually, keeping in mind the relations
\[
	(p_i)^{q-1}\frac{\partial}{\partial \theta^i} 
		=\frac{\partial}{\partial p_i}, \quad i=1,\cdots,n+1
\]
and 
\[
	D_{\frac{\partial}{\partial \theta^i}} 
		\frac{\partial}{\partial \theta^j}=0, 
	\quad i,j=1,\cdots,n+1,
\]
we have
\[
	D_{\frac{\partial}{\partial p_i}} \frac{\partial}{\partial p_j}=
	(q-1) \frac{\delta_{ij}}{p_i} \frac{\partial}{\partial p_j}
	\quad i,j=1,\cdots,n+1.
\]
Using these relations, we can calculate the Gauss and Weingarten formulas 
(Cf. Appendix) for $\partial_i$ defined in (\ref{basis}) and $\xi$ 
as follows:
\begin{eqnarray*}
	D_{\partial_i} \partial_j 
&=& (q-1)\left\{\frac{\delta_{ij}}{p_i}\frac{\partial}{\partial p_i} 
	+ \frac{1}{p_{n+1}}\frac{\partial}{\partial p_{n+1}} \right\} \\
\displaystyle
&=& \sum_{k=1}^{n}\Gamma^{k}_{ij} \partial_k + h_{ij} \xi, 
\\
	D_{\partial_i} \xi &=& q(q-1) \left\{ \frac{\partial}{\partial p_i}
	- \frac{\partial}{\partial p_{n+1}} \right\} \\
	&=& -\sum_{j=1}^{n} s^j_i \partial_j.
\end{eqnarray*}
Here, we have used the expression:
\[
	\xi=-(1-q) \sum_{i=1}^{n+1} p_i \frac{\partial}{\partial p_i},
\]
which is equivalent to (\ref{trans}).
Then, we see that the calculated $h_{ij}$ and $\Gamma^{k}_{ij}$ respectively 
coincide with (\ref{Rmetric}) and (\ref{alpha}), i.e., 
it holds that $h_{ij}=g_{ij}$ and $\Gamma^{k}_{ij}=\Gamma^{(\alpha)k}_{ij}$.
Thus, the realized manifold coincides with $({\cal S}^n,g,\nabla^{(\alpha)})$.
Further, the affine shape operator $S=(s_i^j)$ and 
transversal connection form $\tau$ are, respectively, 
\[
	s_i^j=(1-q)q \delta_i^j, \quad \tau=0.
\]
By the property F3) in the appendix, 
these two relations show that the realized manifold 
$({\cal S}^n,g,\nabla^{(\alpha)})$ 
is a statistical manifold with constant curvature $\kappa=q(1-q)$.

\begin{figure}[bhtp]
\begin{center}
\resizebox{0.75\columnwidth}{!}{%
  \includegraphics{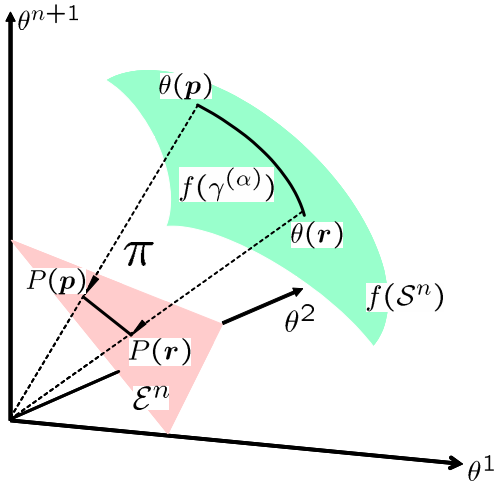}
}
\end{center}
\caption{Projective transformation $\pi$ from $f({\cal S}^n)$ to 
${\cal E}^n$ and the escort distribution $P$.}
\label{fig1}
\end{figure}

This viewpoint clarifies the relation with the 
{\em escort probability}.
Using the coordinates $(\theta^i)$, the escort probability $P_i$
is expressed \cite{BS} by
\[
 P_i(\bm{p}):=\frac{(p_i)^q}{\sum_{i=1}^{n+1} (p_i)^q}
	=\frac{\theta^i(\bm{p})}{\sum_{i=1}^{n+1} \theta^i(\bm{p})}
	\quad i=1,\cdots,n+1.
\] 
Hence, the set of escort probability distributions 
$\bm{P}=(P_i)$ with positive $P_i$, is nothing but the probability simplex 
in the ambient space ${\bf R}^{n+1}_+$.
We denote this set by $\mathcal{E}^n$.
Recall, on the other hand, that the immersion $f(\mathcal{S}^n)$ 
is represented as 
a level surface of $\psi$ in ${\bf R}^{n+1}_+$ (See Figure \ref{fig1}).
Thus, for each $(p_i)$, we can define a projection 
$\pi$ from $f({\cal S}^n)$ to 
${\cal E}^n$ by
\[
	\pi: f({\cal S}^n) \ni \bm{\theta}=(\theta^i) \mapsto 
	\bm{P}=(P_i)=(\lambda \theta^i) 
	\in {\cal E}^n,
\]
where
\[
	\lambda:=\frac{1}{\sum_{i=1}^{n+1} \theta^i}.
\]

Now we consider geometric structure of ${\cal E}^n$ in order to derive 
interesting properties of the escort probabilities.
Since $\mathcal{E}^n$ is contained in a hyperplane of 
${\bf R}^{n+1}$, 
it would be natural to use the flat 
connection\footnote{It corresponds to the 
{\em mixture connection} \cite{AN} for 
the escort distributions $(P_i)$ in the terminology of information 
geometry.} induced from $D$.
We use, for the brevity, the same symbol $D$ for the induced connection 
on ${\cal E}^n$.
Then note that any straight line segment on ${\cal E}^n$ is a geodesic of 
$({\cal E}^n,D)$. 

Let ${\cal T}$ be the 2-dimensional sector defined by
\[
	{\cal T}:=\left\{ \bm{\theta} \left| 
	\bm{\theta}= \beta_1 \bm{\theta}(\bm{p}) 
		+ \beta_2 \bm{\theta}(\bm{r}), \;
	\beta_1 \ge 0,\beta_2 \ge 0 \right. \right\}.
\]
Then it is known \cite[p.~44]{NS} that the $\nabla^{(\alpha)}$-geodesic curve 
$\gamma^{(\alpha)}$ 
connecting $\bm{p}$ and $\bm{r}$ is represented on $f({\cal S}^n)$ by 
\begin{equation}
	f(\gamma^{(\alpha)}) = f({\cal S}^n) \cap {\cal T}.
\label{sector}
\end{equation}

On the other hand, ${\cal T}$ also includes points 
$\bm{P}(\bm{p})$ and $\bm{P}(\bm{r})$ by the definition of the escort probabilities.
The intersection ${\cal E}^n \cap {\cal T}$ is a straight line segment, 
i.e, a geodesic of $({\cal E},D)$, connecting $\bm{P}(\bm{p})$ and 
$\bm{P}(\bm{r})$ (Figure \ref{fig1}).

Thus, $\pi \circ f$ maps every $\nabla^{(\alpha)}$-geodesic curve on 
$({\cal S}^n,\nabla^{(\alpha)})$ to a geodesic (line segment) 
on $({\cal E}^n,D)$.
In this sense, $\pi \circ f$ is a {\em projective transformation} 
from $({\cal S}^n,\nabla^{(\alpha)})$ to a flat manifold $({\cal E}^n,D)$.
The above observation is summarized as follows: 
\begin{proposition}
\label{prop_esc}
The escort distribution $(P_i)$ is geometrically interpreted as 
a normalized affine coordinate system of the manifold ${\cal E}^n$ with 
the flat connection $D$, which is 
projectively transformed by $\pi \circ f$ 
from the probability simplex ${\cal S}^n$ 
with the connection $\nabla^{(\alpha)}$.
\end{proposition}

There are, {\em at least}, two possibilities A) and B) to introduce geometric structure 
to the flat manifold $({\cal E}^n,D)$ equipped with the Legendre duality.

\noindent
A) 
The one is to consider the Kullback-Leibler divergence, i.e., 
$-1$-divergence, for two 
escort distributions $\bm{P}=(P_i)$ and $\bm{P}'=(P'_i)$ on ${\cal E}^n$:
\[
	D^{(-1)}(\bm{P},\bm{P}')=\sum_{i=1}^{n+1} P_i \ln \frac{P_i}{P'_i}.
\]
Then ${\cal E}^n$ is the well-developed dually flat 
statistical manifold \cite{Amari,AN} 
with the Fisher information matrix as a Riemannian metric.
The obtained flat connection $D^*$ on ${\cal E}^n$ 
dual to $D$ is called the {\em exponential connection}.
This structure is helpful when we apply the standard technique 
of statistical physics to the escort distributions (, e.g., \cite{BS}) 
and translate to the usual distributions.

\noindent
B)
The other possibility is to induce from the geometry of the ambient space 
${\bf R}^{n+1}_+$ by regarding ${\cal E}^n$ as its submanifold.
We do not describe the detail (See the note below), but 
we only show the corresponding divergence on ${\cal E}^n$.

Let $\varphi(\bm{\eta})$ be 
the Legendre transform of $\psi(\bm{\theta})$, 
i.e., 
\begin{eqnarray}
 \varphi(\eta)&=&\sup_{\bms{\theta} \in {\bf R}^{n+1}_+} 
	\left\{ \sum_{i=1}^{n+1} \theta^i \eta_i 
	- \psi(\bm{\theta})\right\} \nonumber \\
&=&  \frac{2}{1-\alpha} \sum_{i=1}^{n+1}
 \left(\frac{1+\alpha}{2} \eta_i \right)^{2/(1+\alpha)}, \;
\eta_i := \frac{\partial \psi}{\partial \theta^i}.
\label{dualpotential}
\end{eqnarray}
Construct the {\em canonical divergence} ${\cal D}$ \cite{Amari,AN}
on ${\bf R}^{n+1}_+ \times {\bf R}^{n+1}_+$ using 
$\psi$ and the dual parameters $\eta_i$, then for two points $\bm{\theta}$ and 
$\bm{\theta}'$ in ${\bf R}^{n+1}_+$ we have 
\begin{eqnarray}
  {\cal D}(\bm{\theta},\bm{\theta}') &:=&  \psi(\bm{\theta}) + 
	\varphi(\bm{\eta}') - \sum_{i=1}^{n+1} \theta^i \eta'_i \nonumber \\
	&=&\psi(\bm{\theta})-\psi(\bm{\theta}')
		- \sum_{i=1}^{n+1} \eta'_i (\theta^i-\theta'^i).
\label{alphadiv} 
\end{eqnarray}
The divergence ${\cal D}$ reproduces the $\alpha$-divergence (\ref{alphaD}) 
on $f({\cal S}^n) \times f({\cal S}^n)$ 
due to the relations (\ref{immersion}), (\ref{theta}) and
\begin{equation}
\eta_i(\bm{p}) = L^{(-\alpha)}(p_i)
=\frac{1}{1-q}(p_i)^{1-q},
 \; i=1,\cdots,n+1.
\label{eta2}
\end{equation}
Simultaneously, it induces the following natural divergence for two 
escort distributions $\bm{P}=(P_i)$ and $\bm{P}'=(P'_i)$ on ${\cal E}^n$:
\begin{equation}
  {\cal D}(\bm{P},\bm{P}') =  \psi(\bm{P}) - \psi(\bm{P}') 
	- \sum_{i=1}^{n+1} (P_i - P'_i )P'^*_i,
\label{esc_div} 
\end{equation}
where $P^*_i$ is defined by
\[
	P^*_i:=\frac{1}{1-q} (qP_i)^{(1-q)/q}.
\]

\noindent
{\bf Note:}
The canonical divergence (\ref{alphadiv}) defines the 
flat statistical manifold structure $({\bf R}^{n+1}_+,\tilde g, D)$ 
with the Riemannian metric $\tilde g$:
\begin{eqnarray}
	\tilde g_{ij}
	&:=& -\left. \frac{\partial}{\partial \theta^i}
	\frac{\partial}{\partial \theta'^j}{\cal D}(\bm{\theta},\bm{\theta}') 
	\right|_{\bms{\theta}=\bms{\theta}'} \nonumber \\
	&=& \frac{\partial \psi}{\partial \theta^i \partial \theta^j}
	= (q\theta^i)^{(1-2q)/q}\delta_{ij}.
\label{Hess}
\end{eqnarray}
and the flat connection $D$.
Further, it induces a geometry on 
${\cal S}^n$ via  $f({\cal S}^n)$ as the statistical 
submanifold of ${\bf R}^{n+1}_+$.
This is the essentially same way to \cite{Ohara07} 
or section \ref{shortrev}.
We have seen in this section that the induced geometry coincides 
with the one defined by the affine immersion $(f,\xi)$.
This is due to the 
special choice of the transversal vector $\xi$ satisfying 
(\ref{scale}) and (\ref{normal}). 
See for detail \cite{HS,Shima,UOF1}.

The manifold $({\bf R}^{n+1}_+,\tilde g, D)$
also induces geometry on the submanifold ${\cal E}^n$.
It can be proved to be a flat statistical manifold and its 
corresponding divergence is given in (\ref{esc_div}).
The induced geometry is known to be not only projectively  
but also conformally transformed by $\pi \circ f$ 
from $({\cal S}^n,g,\nabla^{(\alpha)})$, 
which directly follow from the concept of $-1$-conformal equivalence 
\cite{Kurose2}.

\section{Generalization via centro-affine immersion}
\label{GCE}
In this section we discuss a possible generalization of entropy, 
relative entropy (divergence), escort probability 
keeping  statistical manifold structure of ${\cal S}^n$. 
The key idea is by applying the method of centro-affine immersion 
observed in the previous section.

Consider a smooth representing function $s=L(t)$ defined on $t \ge 0$ 
satisfying the following assumptions:
\begin{description}
\item[A1:] strictly increasing,
\item[A2:] $L(0)=0$,
\item[A3:] $d^2 L/dt^2 <0$ for all $t \ge 0$.
\end{description} 
These assumptions ensure that the strictly increasing 
inverse function $L^{-1}$ 
exists and meets conditions: $L^{-1}(0)=0$ and 
$d^2 L^{-1}/ds^2 >0$ for 
all $s \ge 0$.
Hence, $L^{-1}$ is a convex function.

Let $f$ be an affine immersion defined by (\ref{immersion}), then 
${\cal S}^n$ is immersed in ${\bf R}^{n+1}_+$ as a level surface 
of $\psi^{(L)}$:
\[
	f({\cal S}^n):=\left\{ \bm{\theta} \left| 
		\sum_{i=1}^{n+1} \psi^{(L)}(\bm{\theta}) 
		=c, \quad \exists c >0 \right. \right\}.
\]
Here, $\psi^{(L)}$ is a convex function defined by
\[
	\psi^{(L)}(\bm{\theta}):=c \sum_{i=1}^{n+1} L^{-1}(\theta^i),
\]
which has a positive definite Hessian from the assumptions.
The level $c$ is arbitrary.
For the case of the $\alpha$-geometry we have used $L=L^{(\alpha)}$ 
and $c=2/(1+\alpha)$.

We take a transversal vector $\xi$ so that the affine immersion 
$(f,\xi)$ is centro-affine with a scaling constant $\tilde \kappa$:
\[
	\xi= \sum_{i=1}^{n+1}\xi^i 
		\frac{\partial}{\partial \theta^i},
	\quad \xi^i=-\tilde \kappa \theta^i
\]
where $\tilde \kappa$ is an arbitrary constant.
Since $f({\cal S}^n)$ is a strongly convex surface, $\xi$ defined above is 
transversal to $f({\cal S}^n)$.
Hence, according to F3) in the appendix, we find that $(f,\xi)$ realizes 
a statistical manifold $({\cal S}^n,h^{(L)},\nabla^{(L)})$ 
with constant curvature $\tilde \kappa$.

Instead of the canonical divergence, we invoke the 
{\em geometric divergence} \cite{Kurose2}:
\begin{equation}
	D^{(L)}(\bm{p},\bm{r}):=
		- \sum_{i=1}^{n+1} \nu_i(\bm{r}) 
			(\theta^i(\bm{p})-\theta^i(\bm{r})), 
\label{geom_div}
\end{equation}
where $\nu=\sum_{i=1}^{n+1} \nu_i d\theta^i \in ({\bf R}^{n+1})^*$ 
is called the {\em conormal vector} \cite{NS} defined by
\begin{equation}
	\nu_i:=\frac{1}{\Lambda} 
	\frac{\partial \psi^{(L)}}{\partial \theta^i}, 
	\; i=1,\cdots ,n+1,
\label{special}
\end{equation}
where
\[
	\quad \Lambda:= d\psi^{(L)}(\xi)=\sum_{i=1}^{n+1} 
	\frac{\partial \psi^{(L)}}{\partial \theta^i} \xi^i .
\]
Here, $({\bf R}^{n+1})^*$ denotes the dual space of ${\bf R}^{n+1}$ 
and $\nu_i$ plays a similar role of the dual parameter $\eta_i$ in the case of 
the $\alpha$-geometry.

It is known \cite{Kurose2} 
that the analogous statement to Proposition \ref{prop_mP} 
holds for $D^{(L)}$, i.e., 
for a right triangle on ${\cal S}^n$ with respect to the realized 
Riemannian metric $h^{(L)}$ and dual connections $\nabla^{(L)}$ and 
$\nabla^{(L)*}$, it holds that 
\begin{eqnarray}
	D^{(L)}(\bm{p},\bm{r}) &=& 
 D^{(L)}(\bm{p},\bm{q}) + D^{(L)}(\bm{q},\bm{r}) \nonumber \\
&& \; -\tilde \kappa D^{(L)}(\bm{p},\bm{q})D^{(L)}(\bm{q},\bm{r}).
\label{mPm}
\end{eqnarray}
Using the uniform distribution $\bm{u}$, 
we can define an associated entropy $S^{(L)}$ 
with the divergence $D^{(L)}$ by
\begin{equation}
	S^{(L)}(\bm{p}):=M-D^{(L)}(\bm{p},\bm{u})
\end{equation}
where
\[
	M:=\max_{\bms{p} \in \bar{{\cal S}}^n} D^{(L)}(\bm{p},\bm{u}).
\]
Then $S^{(L)}$ is confirmed to meet positivity, convexity, 
continuity and take the maximum at $\bm{p}=\bm{u}$.
The modified Pythagorean relation (\ref{mPm}) is essentially important 
to investigate the minimization of the generalized divergence $D^{(L)}$ or 
maximization of the generalized entropy $S^{(L)}$, as is in the case of 
$K_q$ or $S_q$ \cite{Ohara07}.

The generalized escort probability (see also \cite{BS}) 
is similarly defined by
\[
	P_i^{(L)}:=\frac{L(p_i)}{\sum_{i=1}^{n+1} L(p_i)}.
\]
The projectivity and the property in Proposition \ref{prop_esc} 
are inherited to this generalization.

However, unlike the case of the previous section, 
the divergence defined on ${\bf R}^{n+1}_+$ by (\ref{alphadiv}) 
with $\psi^{(L)}$ instead of $\psi$ does {\em not} generally induce 
$D^{(L)}$ in (\ref{geom_div}) on ${\cal S}^n$. 
By comparing (\ref{alphadiv}) and (\ref{geom_div}), we see that 
the property holds if and only if $\nu_i$ coincides with the 
dual parameter of $\theta^i$, i.e., 
\[
	\nu_i=\frac{\partial \psi^{(L)}}{\partial \theta^i}, 
	\quad i=1,\cdots,n+1.
\]
From (\ref{special}) it is equivalent to the condition that $\Lambda=1$, 
i.e., 
\[
	d\psi^{(L)}(\xi) =\tilde \kappa \sum_{i=1}^{n+1} 
	\frac{\partial \psi^{(L)}}{\partial \theta^i}\theta^i =1
\]
holds on $f({\cal S}^n)$.
Thus, the property is specific to the case $L$ is a power function like 
(\ref{theta}).

\section{Application (1): Alpha-convexity and autoparallelism of 
submanifolds constrained by expectations}
\label{app1}
In this section, we apply the geometry 
$({\cal S}^n,g,\nabla^{(\alpha)})$ to discuss convexity of 
submanifolds constrained by the modified expectations (averages).
Convexity of the submanifold is crucial in considering the 
maximization of $S_q$. 
Here we restrict to the case that $q>0, q\not=1$.
The results can be generalized to 
$({\cal S}^n,h^{(L)},\nabla^{(L)})$ via the analogous arguments.

The subset ${\cal A}$ in ${\cal S}^n$ is said 
{\em $\nabla^{(\alpha)}$-autoparallel}, 
if it holds that 
\[
	f({\cal A})=f({\cal S}^n) \cap {\cal T}
\]
for a certain open convex set ${\cal T}$ contained in 
a linear subspace of ${\bf R}^{n+1}$ with respect to 
the $\theta$-coordinate system. 
The characterization of the $\nabla^{(\alpha)}$-geodesic 
curve of 
(\ref{sector}) is the special case where the dimension of 
the linear subspace is two.
The subset ${\cal C}$ in ${\cal S}^n$ is said {\em $\nabla^{(\alpha)}$-convex} 
if the $\alpha$-geodesic curve connecting arbitrary two points on ${\cal C}$ 
is contained in ${\cal C}$.
Note that 
${\cal A}$ is $\nabla^{(\alpha)}$-convex if it is 
$\nabla^{(\alpha)}$-autoparallel.

The $\nabla^{(\alpha)}$-convexity or 
$\nabla^{(\alpha)}$-autoparallelism are 
important in studying the {\em Tsallis relative entropy minimization} 
from a viewpoint of the optimization.
It is because the modified Pythagorean theorem guarantees 
the uniqueness of the minimizing distribution 
on the $\nabla^{(\alpha)}$-convex set. 
In this case, the minimizing distribution is characterized by 
the so-called {\em $-\alpha$-projection} \cite{Amari,AN,Ohara07}.

Let $A_i$ be the quantity assigned to the $i$-th microstate and
consider the linear, $q$- and $q$-{\em normalized expectations} 
\cite{TSP}, respectively defined by
\begin{eqnarray*}
 &  \langle A \rangle  := \sum_{i=1}^{n+1} p_i A_i, \quad 
   \langle A \rangle _q := \sum_{i=1}^{n+1} (p_i)^q A_i, & \\
& \displaystyle \langle \langle A \rangle \rangle _q := 
\frac{\sum_{i=1}^{n+1} (p_i)^q A_i}{\sum_{i=1}^{n+1} (p_i)^q}. &
\end{eqnarray*}


For a prescribed value $\bar A$, define the constrained manifolds 
in $\mathcal{S}^n$ by
\begin{eqnarray*}
 & \mathcal{H}^+:=\mathcal{S}^n \cap \{\bm{p} | \langle A \rangle  \ge \bar
 A \}, \quad
 \mathcal{H}_q^+:=\mathcal{S}^n \cap \{\bm{p} | \langle A \rangle _q 
\ge \bar A \}, & \\
& \widetilde {\mathcal{H}}_{q}^+ := \mathcal{S}^n \cap \{\bm{p} |
\langle \langle A \rangle \rangle_q \ge \bar A \}. &
\end{eqnarray*}
Similarly, $ \mathcal{H}^-, \mathcal{H}_q^-,
\widetilde{\mathcal{H}}_q^-$, and the boundaries 
$\mathcal{H}, \mathcal{H}_q, \widetilde{\mathcal{H}}_q$ 
can be defined by replacing the inequality 
symbols by the reverse and the equality ones, respectively.

Since the constraints for $\mathcal{H}_q^\pm,
\widetilde{\mathcal{H}}_q^\pm, \mathcal{H}_q$ or 
$\widetilde{\mathcal{H}}_q$ are nonlinear with respect to $\bm{p}$, 
the corresponding equilibrium distributions for the Tsallis relative entropy 
$K_q(\cdot,\bm{r})$ (or $\alpha$-divergence $D^{(\alpha)}(\cdot,\bm{r})$) 
are not necessarily unique nor the minimizers for them, while 
they are convex in the usual sense 
with respect to $\bm{p}$ when $q>0$ and $q \not= 1$.

Let the constrained subsets be nonempty. 
Then they have the following properties:
\begin{description}
 \item[1)] $\mathcal{H}^-$ (resp. $\mathcal{H}^+$) is 
$\nabla^{(\alpha)}$-convex if
	   $0<q<1$ ($q>1$) and $A_i > 0$ for all $i$,
 \item[2)] $\mathcal{H}_q^+$ $(\mathcal{H}_q^-)$ is 
$\nabla^{(\alpha)}$-convex if
	   $0<q<1$ $(1<q)$,
 \item[3)] Let $0<q<1$ ($1<q$) and $0< A_i$ ($0>A_i$) for all $i$. 
Then $\mathcal{H}_q$ is a $D^{(\alpha)}$-sphere, i.e., 
\[
	   \mathcal{H}_q= \mathcal{S}^n \cap \{\bm{p} | 
	   D^{(\alpha)}(\bm{p},\bm{r})=d \}
\] 
for some $\bm{r} \in {\cal S}^{n}$ and $d \in \mathbf{R}$,
 \item[4)] Both $\widetilde{\mathcal{H}}_q^\pm $ are 
$\nabla^{(\alpha)}$-convex,
 \item[5)] $\widetilde{\mathcal{H}}_q $ is a $\nabla^{(\alpha)}$-autoparallel 
	   submanifold in $\mathcal{S}^n$.
\end{description}

All of the above statements follow from the 
standard convexity argument with respect to the $\theta$-coordinate 
based on the definition of the $\alpha$-convexity.
In particular, note that constraints $\langle A \rangle _q =
\bar A$ and $ \langle \langle A \rangle \rangle _q = \bar A$ are 
respectively characterized by linear constraints in $\theta$, i.e., 
\[
q \sum_{i=1}^{n+1} A_i \theta^i = \bar A, \qquad \sum_{i=1}^{n+1}
(A_i - \bar A) \theta^i = 0.  
\]
For 3), recall that the $\alpha$-divergence can be alternatively 
expressed, using (\ref{eta2}), by
\begin{equation}
 D^{(\alpha)}(\bm{p},\bm{r}) = 
\frac{1}{q(1-q)}-\sum_{i=1}^{n+1} \theta^i(\bm{p}) \eta_i(\bm{r}).
\label{alphadiv2}
\end{equation}
Then we can verify the statement by setting 
\[
	\eta_i(\bm{r}):=\epsilon^{q-1}A_i, \quad
	d:=\frac{1}{q}\left( \frac{1}{1-q} - \epsilon^{q-1} \bar{A} \right),
\]
where
\[
	\epsilon:=\sum_{i=1}^{n+1} ((1-q)A_i)^{1/(1-q)}.
\]

For the detail of the Tsallis relative entropy minimization with 
the constraints of $\widetilde{\mathcal{H}}_q$, see \cite{Ohara07}.

\section{Application (2): Gradient flow of the alpha-divergence}
\label{GrFl}
As another application we investigate the gradient flow of 
the $\alpha$-divergence. 
Let $X=\sum_{i=1}^n X^i \partial_i$ be the gradient vector on 
$\mathcal{S}^n$ minimizing $D^{(\alpha)}(\bm{p},\bm{r})$ for 
a given distribution $\bm{r}$, where $\partial_i$ is defined 
by (\ref{basis}). 
Then the component $X^i$ is expressed by 
\begin{equation}
 X^i = -\sum_{j=1}^n g^{ij} \partial_j 
	D^{(\alpha)}(\bm{p},\bm{r}), \quad i=1,\cdots,n,  
\label{gradient}
\end{equation}
where $(g^{ij})$ is the inverse matrix of the Riemannian metric 
$g=(g_{ij})$ in (\ref{Rmetric}).

Since $D^{(\alpha)}(\bm{p},\bm{r})$ is strongly convex with respect to 
$\bm{p}$ in ${\cal S}^n$, the flow converges to $\bm{r}$.
Further, the constant curvature property of 
$(\mathcal{S}^n,g,\nabla^{(\alpha)})$ gives the restriction for 
the behavior of the gradient flow as follows.
For dually flat case, the corresponding results are found 
in \cite{FA}.

\begin{proposition}(constants of motion)
\label{prop_com}
 The trajectory of 
the gradient flow $\bm{p}(t)$ of $D^{(\alpha)}(\bm{p},\bm{r})$ on 
$(\mathcal{S}^n,g,\nabla^{(\alpha)})$ with an initial point $\bm{p}_0$ 
coincides with the $\nabla^{(-\alpha)}$-geodesic curve 
connecting $\bm{p}_0$ and $\bm{r}$.
Further, let $\bm{\theta}_l=(\theta^i_l) \in {\bf R}^{n+1}, 
l=1,\cdots,n-1$ be the set of linearly independent vectors satisfying
\[
	\sum_{i=1}^{n+1}\theta_l^i \eta_i(\bm{p}_0)=0, \quad 
	\sum_{i=1}^{n+1}\theta_l^i \eta_i(\bm{r})=0, 
\quad l=1,\cdots,n-1.
\]
Then, the quantities $C_l$ defined by
\[
	C_l:=\sum_{i=1}^{n+1}\theta_l^i \eta_i(\bm{p}(t)), 
	\quad l=1, \cdots, n-1
\]
are the $n-1$ independent constants of motion for the gradient flow.
\end{proposition}
See appendix B for the proof. 

\section{Generalized exponential family and U-geometry}
\label{GEU}

Now we discuss the Legendre duality in the case of iii) in section \ref{sec2}, 
following \cite{Eguchi,MTKE,Naudts02,Naudts041,Naudts043}.
For a fixed strictly increasing and positive function $\phi(s)$ on 
$(0,\infty)$, define 
the {\em generalized logarithmic function} as follows:
\begin{equation}
\ln_\phi(t):=\int_1^t \frac{1}{\phi(s)}ds, \quad t>0.  \;
\label{glog}
\end{equation}
The {\em generalized exponential function} denoted by $\exp_\phi$ is defined as 
the inverse function of $\ln_\phi$.

Define a convex function $F_\phi(s)$ for $s>0$ by
\begin{eqnarray}
 F_\phi(s):=\int_1^s \ln_\phi t dt, \quad \lim_{s \rightarrow 0_+}F_\phi(s)<+\infty \hbox{ :assumed}.
\label{gl}
\end{eqnarray}
For probability density functions $p(x)$ and $r(x)$,
introduce a {\em generalized entropy} functional defined by
\begin{equation}
 {\cal I}_\phi[p]:= 
 \int -F_\phi(p(x))+(1-p(x))F_\phi(0) dx,
\label{g_entropy}
\end{equation}
and the {\em Bregman divergence} defined by 
\begin{equation}
 {\cal D}_\phi[p \| r]:=\int U_\phi(\ln_\phi r)-U_\phi(\ln_\phi p)
	-p(\ln_\phi r-\ln_\phi p) dx,
\label{DIVdual}
\end{equation}
where the function $U_\phi$ is the Legendre conjugate of $F_\phi$ 
defined by
\begin{equation}
	U_\phi(t):=t \exp_\phi t -F_\phi(\exp_\phi t).
\label{Leg}
\end{equation}


Let us consider the following finite dimensional statistical model 
called {\em the generalized exponential family} \cite{GD}
($\phi$-exponential family \cite{Naudts043} 
or $U$-statistical model \cite{Eguchi}), 
which is defined by
\[
 \mathcal{M}_\phi= 
\{p_\theta(x)=\exp_\phi(\theta^T h(x)-\psi_\phi(\theta)) | \theta \in 
\Omega \subset {\bf R}^d \} 
\]
where $h(x)=(h_i(x)), i=1,\cdots,d$ is a certain vector-valued function 
and $\psi_\phi(\theta)$ is a normalizing factor of $p_\theta(x)$.

Introduce the following {\em potential function}:
\[
 \Psi_\phi(\theta) 
:= \int U_\phi(\ln_\phi p_\theta)+(1-p_\theta)F_\phi(0) dx 
+\psi_\phi(\theta).
\]
It follows from the relation $\exp_\phi=U'_\phi$ that
\begin{equation}
 \eta_i(\theta):=\partial_i \Psi_\phi(\theta)= \int h_i(x)p_\theta(x)dx
=\mathbf{E}_{p_\theta}[h_i(x)],
\label{expectation}
\end{equation}
where $\partial_i:=\partial/\partial \theta^i$ and 
we denote by ${\bf E}_p[\cdot]$ the expectation operator 
for the density $p$.
Then, the Hesse matrix of $\Psi_\phi(\theta)$ is expressed by
\begin{equation}
 \partial_i \partial_j \Psi_\phi(\theta)
= \int \tilde h_i(x) 
	\exp'_\phi(\theta^T h(x)-\psi_\phi(\theta)) 
	\tilde h_j(x) dx, 
\label{Hesse}
\end{equation}
where $\tilde h_i(x):=h_i(x)-\partial_i \psi_\phi(\theta)$.
We see that it is positive semidefinite because $\exp'_\phi$ is positive, 
and hence, $\Psi_\phi$ is a convex function of $\theta$.
In the sequel, we assume that 
$(\partial_i \partial_j \Psi_\phi)=(\partial \eta_j/\partial \theta^i)$ 
is positive definite for $\forall \theta \in \Omega$.
Hence, $\eta=(\eta_i)$ is locally bijective to $(\theta^i)$ and 
we call $\eta=(\eta_i)$ the {\em expectation coordinate system} 
for ${\cal M}_\phi$.
By the relation (\ref{expectation}) the Legendre conjugate of 
$\Psi_\phi(\theta)$ is the sign-reversed generalized entropy 
of $p_\theta \in {\cal M}_\phi$, i.e,
\begin{equation}
 \Psi^*_\phi(\eta) = \theta^T \eta - \Psi_\phi(\theta) 
= -{\cal I}_\phi[p_\theta].
\label{Massieu}
\end{equation}
Hence, $\Psi_\phi(\theta)$ can be physically interpreted as the 
{\em generalized Massieu potential} \cite{Callen,WS05} 
and our Riemannian metric $(\partial_i \partial_j \Psi_\phi)=(\partial \eta_j/\partial \theta^i)$ introduced below is regarded as a {\em susceptance} matrix.


The U-geometry \cite{Eguchi} is introduced as follows:
As a {\em Riemannian metric} $g=(g_{ij})$ on ${\cal M}_\phi$, 
which is an inner product for tangent vectors, 
we use the Hesse matrix of $\Psi_\phi$.
Note that we can alternatively express (\ref{Hesse}) as
\[
 g_{ij}(\theta)=g(\partial_i,\partial_j):= \partial_i \partial_j \Psi_\phi
=\int \partial_i p_\theta \partial_j
 \ln_\phi p_\theta dx.
\]
Further we define a generalized version of the {\em mixture connection} 
$\nabla^{({\rm m})}$ and {\em exponential connection} 
$\nabla^{({\rm e})}$  
by their components 
\[
 \Gamma^{({\rm m})}_{ij,k} (\theta) 
	=g(\nabla^{({\rm m})}_{\partial_i} \partial_j,\partial_k)
	:= \int \partial_i \partial_j p_\theta 
		\partial_k \ln_\phi p_\theta dx,
\]
\begin{equation}
 \Gamma^{({\rm e})}_{ij,k} (\theta) 
	=g(\nabla^{({\rm e})}_{\partial_i} \partial_j,\partial_k)
	:= \int \partial_k  p_\theta
		\partial_i \partial_j \ln_\phi p_\theta dx.
\label{dual_connex} 
\end{equation}
Then the duality relation of the connections \cite{Amari,AN} holds, i.e.,
$
	\partial_i g_{jk}=\Gamma^{({\rm m})}_{ij,k}+\Gamma^{({\rm e})}_{ik,j}
$.
Further, ${\cal M}_\phi$ can be proved to be flat with respect to both 
$\nabla^{({\rm m})}$ and $\nabla^{({\rm e})}$.
Thus, we have obtained {\em dually flat} \cite{AN} structure 
$(g,\nabla^{({\rm m})}, \nabla^{({\rm e})})$  
on ${\cal M}_\phi$ defined by the derivatives of $\Psi_\phi$.

\begin{proposition}
\label{prop_geod}
Let ${\cal C}$ be a one-dimensional 
submanifold on ${\cal M}_\phi$.
If ${\cal C}$ is expressed as a 
straight line in the coordinates $\theta$,
then ${\cal C}$ coincides with a $\nabla^{({\rm e})}$-geodesic 
({\em e-geodesic}, in short) curve.
If ${\cal C}$ is expressed as a 
straight line in the coordinates $\eta$,
then ${\cal C}$ coincides with a 
$\nabla^{({\rm m})}$-geodesic ({\em m-geodesic}) curve.
\end{proposition}

\begin{definition}
Let $p(x)$ be a given density. 
If there exists the minimizing density function $\hat p_\theta(x)$ 
for the variational problem 
$\min_{p_\theta \in {\cal M}_\phi} {\cal D}_\phi[p \| p_\theta]$,
or equivalently, the minimizing parameter $\hat \theta$ for the problem 
$\min_{\theta \in \Omega}{\cal D}_\phi[p \| p_\theta]$
exists, we call $\hat p_\theta(x)=p_{\hat \theta}(x)$ 
the {\em m-projection of $p(x)$ to ${\cal M}_\phi$}.
\end{definition}

\begin{proposition}
\label{prop_proj}
Let $\hat p_\theta \in {\cal M}_\phi$ be the m-projection of $p$.
Then the following properties hold:
\begin{description}
\item[i)] The expectation of $h(x)$ is conserved by the m-projection, i.e., 
${\bf E}_p[h(x)]={\bf E}_{\hat p_\theta}[h(x)]$,
\item[ii)] The following triangular equality holds:
$
	{\cal D}_\phi[p \| p_\theta]={\cal D}_\phi[p \| \hat p_\theta] 
	+ {\cal D}_\phi[\hat p_\theta \|p_\theta]
$ for all $p_\theta \in {\cal M_\phi}$.
\end{description}
\end{proposition}

\begin{remark}
\label{rem_proj}
From the statement i) the m-projection $\hat p_\theta$ 
is characterized as the density in 
${\cal M}_\phi$ with the equal expectation of $h(x)$ to that for $p$.
Note that the following relation:
\begin{eqnarray*}
	{\cal D}_\phi[p \| \hat p_{\theta}]&=&
	\Psi_\phi(\hat \theta)-{\cal I}_\phi[p]
	-\hat \theta^T {\bf E}_p[h(x)] \\
	&=& \Psi_\phi(\hat \theta)-\hat \theta^T \hat \eta-{\cal I}_\phi[p]
	= {\cal I}_\phi[\hat p_\theta]-{\cal I}_\phi[p] \ge 0.
\end{eqnarray*}
Thus, $\hat p_\theta$ achieves the maximum entropy among densities 
with the equal expectation of $h(x)$.
\end{remark}

\section{Application (3) : Nonlinear diffusion equation}
\label{NDE}
Let $u(x,t)$ and $p(x,\tau)$ 
on ${\bf R}^n \times {\bf R}_+$ be, respectively, the solutions of
the following nonlinear diffusion equation, which is called the
{\em porous medium equation} (PME):
\begin{equation}
 \frac{\partial u}{\partial t}=\Delta u^m, \quad m>1
\label{PME}
\end{equation}
with nonnegative initial data $0 \le u(x,0)=u_0(x) \in L^1(\mathbf{R}^n)$, 
and the associated {\em nonlinear Fokker-Planck equation} (NFPE):
\begin{eqnarray}
   \frac{\partial p}{\partial \tau} = \nabla \cdot 
	\left( \beta x p + D \nabla p^m \right), \quad \beta >0 
\label{DNFPE}
\end{eqnarray}
with nonnegative initial data $0 \le p(x,0)=p_0(x) \in L^1(\mathbf{R}^n)$.
Here, $D$ is a real symmetric positive definite matrix, which represents 
the diffusion coefficients.
As is widely known \cite{Vaz,Toscani} and shown later, 
solutions of the both equations are related with 
a simple transformation.


The PME and NFPE with $m>1$ represent the so-called 
{\em slow diffusion} phenomena, 
which naturally arises in many physical problems including 
percolation of a fluid through porous media and so on.
See for \cite{Mu,Buck,LP,Kath,Bar2} and the references therein.
Hence the behaviors of their solutions have been extensively studied 
in both analytical and thermostatistical aspects in the literature 
\cite{FK,Ar,Newman,Ralston,PlaPla,Shiino01,Frank01,Frank02,CT,Otto,Vaz,Toscani,Bar,Pa}, 
just to name a few.

In this section we demonstrate that several geometric concepts derived with 
generalized entropies in the previous section are useful to investigate 
a new aspects of the behavior of the above equations.
For the proofs of the results see \cite{OW08}.

\subsection{Several geometric properties of the porous medium and the associated Fokker-Planck equation}
\label{sec_main}

Set $\phi(u)=u^q, q>0, q\not=1$, then we have the $q$-logarithmic and
exponential functions \cite{TLSM}:
\begin{eqnarray*}
 \ln_\phi t=\ln_q t &:=& (t^{1-q}-1)/(1-q), \\
 \exp_\phi t=\exp_q t &:=& [1+(1-q)t]_+^{1/(1-q)}.
\end{eqnarray*}
Consider the $q$-Gaussian density function defined by:
\begin{eqnarray}
& f(x;\theta,\Theta)=
\exp_q\left(\theta^Tx +x^T \Theta  x - \psi(\theta, \Theta)\right), &
\label{qGauss} \\
& \quad \theta=(\theta^i) \in {\bf R}^n, 
\Theta=(\theta^{ij}) \in {\bf R}^{n \times n}, \nonumber &
\end{eqnarray}
where 
$\Theta$ is a real symmetric negative definite matrix
and $\psi(\theta,\Theta)$ is a normalizing constant.
We denote by ${\cal M}$ the set of $q$-Gaussian densities, i.e., 
\begin{equation}
 {\cal M}:=\left\{ \left. f(x;\theta, \Theta)
\right| \theta \in {\bf R}^n, \; 0>\Theta =\Theta^T \in {\bf R}^{n \times n} \right\}.
\label{qGaussmodel}
\end{equation}
For this setting, the corresponding generalized entropy and divergence are
\begin{equation}
 {\cal I}[p] 
= \frac{1}{2-q}\int \frac{p(x)^{2-q}-p(x)}{q-1}dx
\label{dualent}
\end{equation}
\begin{equation}
	{\cal D}[p\|r]=\int \frac{r(x)^{2-q}
	-p(x)^{2-q}}{2-q}-p(x)\frac{r(x)^{1-q}-p(x)^{1-q}}{1-q}dx, 
\end{equation}

In the sequel we fix the relation between the exponents of the PME 
and the parameter of $q$-exponential function by $m=2-q$.
Hence, we consider the case $1<m<2$, or equivalently, $0<q<1$.
Since we fix $\phi(u)=u^q$, we omit the subscripts $\phi$ used 
to denote several quantities.
By a suitable linear scaling of $t$ we can consider the problem 
by fixing $\beta$ to an arbitrary constant.
Hence, we fix $\beta$ and 
introduce another constant $\mu$ for notational simplicity as follows:
\[
	\beta:=\frac{1}{n(m-1)+2}, \quad \mu:=n \beta.
\]

For the $q$-Gaussian family ${\cal M}$, we can regard 
$(\theta,\Theta)$ as the canonical coordinates, and the first moment vector 
and second moment matrix $(\eta,H)$ defined by
\[
	\eta:=\int xf(x;\theta,\Theta)dx, \quad 
	H:=\int xx^T f(x;\theta,\Theta)dx,
\]
as the expectation coordinates, respectively.

We assume the $u(x,0)$ and $p(x,0)$ are
nonnegative and integrable function with finite second moments.
When we consider the set of solutions, we restrict their initial masses 
to be normalized to one without loss of generalities.

It is proved that there exists a unique nonnegative weak solution 
if $m>0$ \cite[Theorem 5.1]{Vaz}, and that 
the mass $\int u(x,t)dx$ is invariant for all $t>0$ 
if $m \ge (n-2)/n$ \cite{Vaz}.

First of all, we review how the solutions of the PME and NFPE relate in the 
proposition below.
Because of this fact the properties of the solution of the PME (\ref{PME}) 
are important to investigate those of the NFPE (\ref{DNFPE}) and vise versa.


\begin{proposition}
\label{prop_transfer}
Let $u(x,t)$ be a solution of the PME (\ref{PME}) with initial data 
$
	u(x,0)=u_0(x) \in L^1({\bf R}^n).
$
Define 
\[
	p(z,\tau):=(t+1)^\mu u(x,t), \quad z:=(t+1)^{-\beta}Rx , \; \tau:=\ln (t+1),
\]
then $p(z,\tau)$ is a solution of (\ref{DNFPE}) with $\nabla=\nabla_z$, 
$D=RR^T$ 
and initial data
$
	p(z,0)= u_0(R^{-1}z).
$
\end{proposition}

Next, we find that the equilibrium density for the NFPE is 
on the $q$-Gaussian family ${\cal M}$ via Lyapunov approach.
To analyze the behavior of (\ref{DNFPE}) 
let us define {\em generalized free energy}:
\[
 {\cal F}[p]
	:= \int \frac{\beta}{2m} x^T D^{-1} x p(x) dx -{\cal I} [p] 
\]
This type of functional was first introduced in \cite{Newman,Ralston}.
We have 
\begin{equation}
\frac{d{\cal F}[p(x,\tau)]}{d\tau} 
=- \frac{1}{m}\int p \| \beta R^{-1} x + m p^{m-2}R \nabla p \|^2 dx \le 0.
\label{Lyapunov}
\end{equation}
Thus, the equilibrium density $p_\infty(x)$ is  
determined from (\ref{Lyapunov}) as a $q$-Gaussian:
\begin{equation}
	p_\infty(x)=f(x;0,\Theta_\infty) 
	=\exp_q(x^T \Theta_\infty x-\psi(0, \Theta_\infty)),
\label{eq_density}
\end{equation}
where the canonical parameters are given by
\[
	\theta_\infty=0, \quad \Theta_\infty=-\frac{\beta}{2m} D^{-1}.
\]

Note that we can express the difference of the free energies of $p(x)$ and 
the equilibrium $p_\infty(x) \in {\cal M}$ 
by the divergence:
\begin{eqnarray*}
 {\cal D}[p||p_\infty] &=&\Psi(0,\Theta_\infty)
	-{\cal I}[p] 
	- \Theta_\infty \cdot {\bf E}_p[xx^T]  \\
	&=& {\cal F}[p]-{\cal F}[p_\infty].
\end{eqnarray*}
Thus, the minimization of ${\cal F}[\cdot]$ 
is equivalent to that of ${\cal D}[\cdot \|p_\infty]$.

Finally, we show one of the fundamental properties of the PME 
and NFPE, which is important to state the sequel results in this paper.

\begin{proposition}
The $q$-Gaussian family ${\cal M}$ is an invariant
manifold for both PME and NFPE.
\end{proposition}

\subsection{Trajectories of m-projections}
Let $\eta^{\rm PM}=(\eta^{\rm PM}_i)$ and $H^{\rm PM}=(\eta^{\rm PM}_{ij})$ 
be, respectively, the first moment vector and 
the second moment matrix, i.e., 
\[
	\eta^{\rm PM}_{i}(t):=\mathbf{E}_u[x_i]=\int x_i u(x,t) dx,
	\quad \eta^{\rm PM}_{ij}(t):=\mathbf{E}_u[x_ix_j].
\]
\begin{theorem}
\label{thm1}
Consider solutions of the PME 
with the common initial first and second moments.
Then their m-projections to ${\cal M}$ evolve monotonically along 
with the common m-geodesic curve that starts from the density 
determined by the initial moments.
\end{theorem}
Outline of the proof) 
Differentiating $\eta^{\rm PM}_{ij}$ by $t$, 
we see that the second moments evolves as
\begin{eqnarray*}
 \eta^{\rm PM}_{ij}(t)&=&\eta^{\rm PM}_{ij}(0)+ \delta_{ij} \sigma_u^{\rm PM}(t), \\ 
\sigma_u^{\rm PM}(t)&:=& 2 \int_0^t dt' \int u(x,t')^m dx.
\end{eqnarray*}
Note that $\sigma_u^{\rm PM}(t)$ is positive and monotone increasing on $t>0$.
By similar argument we see that $\dot \eta_i^{\rm PM}=0$, i.e., 
the first moment vector is invariant.
From Proposition \ref{prop_geod} and i) of 
Proposition \ref{prop_proj}, the statement follows.
\qed
\begin{remark}
\label{rem_thm}
i) From the argument for the NFPE, we will see that $\sigma_u^{\rm PM}(t)=O(t^{2 \beta})$ as 
$t \rightarrow \infty$. 
\par \noindent
ii) The theorem implies 
that the trajectories of m-projections on ${\cal M}$ 
for all the PME solutions $u(x,t)$ 
are parallelized in the expectation coordinates, i.e., 
\begin{eqnarray}
	\eta^{\rm PM}(t)&=&\eta^{\rm PM}(0), \\
	H^{\rm PM}(t)&=&H^{\rm PM}(0)+\sigma_u^{\rm PM}(t)I,
\label{sol_mom}
\end{eqnarray}
where $I$ denotes the identity matrix.
See \cite{OW08} for the argument on the constants of motion.
\end{remark}

\begin{figure}[bhtp]
\begin{center}
\resizebox{0.75\columnwidth}{!}{%
  \includegraphics{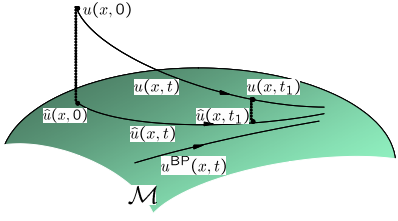}
}
\end{center}
\caption{A solution $u(x,t)$ of the PME, its m-projection $\hat u(x,t)$ 
and the Barenblatt-Pattle solution $u^{\rm BP}(x,t)$ on ${\cal M}$.}
\label{fig2}
\end{figure}

Let $\hat f_0(x) \in {\cal M}$ be the m-projection of the density $f_0(x)$.
Consider two solutions $u_1(x,t)$ and $u_2(x,t)$ of the PME satisfying 
$u_1(x,t_0)=f_0(x)$ and $u_2(x,t_0)=\hat f_0(x)$ for a certain $t_0$.
From the moment conservation property of the m-projection 
stated in Proposition \ref{prop_proj}, 
the second moment matrices $H^{\rm PM}_i(t)$ of $u_i(x,t)$ for $i=1,2$
satisfy $H^{\rm PM}_1(t_0)=H^{\rm PM}_2(t_0)$.
However, their velocities at $t_0$ have the relation:
\begin{eqnarray*}
	\dot H^{\rm PM}_1(t_0)-\dot H^{\rm PM}_2(t_0)
	&=& 2  \int f_0^m(x)-\hat f_0^m(x) dx \; I \\
	&=& 2m(m-1) 
	\left( {\cal I}[\hat f_0]-{\cal I}[f_0] \right)I
\end{eqnarray*}
from (\ref{sol_mom}) and the expression of the generalized entropy 
(\ref{dualent}). 
Using the relation in Remark \ref{rem_proj}, we have the following:
\begin{corollary}
Let $\hat f_0(x) \in {\cal M}$ be the m-projection of a density 
$f_0(x)$ and assume 
that two solutions $u_1(x,t)$ and $u_2(x,t)$ of the PME satisfy the conditions 
$u_1(x,t_0)=f_0(x)$ and $u_2(x,t_0)=\hat f_0(x)$ at $t=t_0$.
Then velocities of their respective second moment matrices at $t_0$ are
related by
\[
	\dot H^{\rm PM}_1(t_0)-\dot H^{\rm PM}_2(t_0)
	=2m(m-1){\cal D}[f_0 \| \hat f_0]I.
\]
\end{corollary}

\begin{figure}[bhtp]
\begin{center}
\resizebox{0.75\columnwidth}{!}{%
  \includegraphics{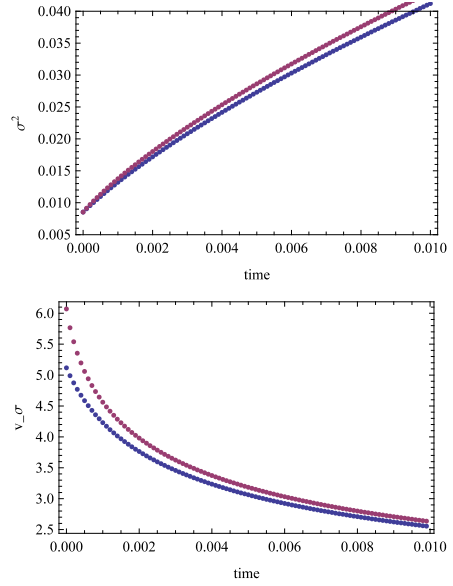}
}
\end{center}
\caption{Evolutions of the second moments of 
$u_1(x,t) \not\in {\cal M}$ 
and $u_2(x,t) \in {\cal M}$ with the same initial moments 
for the one-dimensional PME ($m=1.9$) (Above) 
and the corresponding velocities (Below).}
\label{fig3}
\end{figure}

Thus, the m-projection $\hat u_1(x,t)$ of $u_1(x,t) \not\in {\cal M}$, 
which has the common second moment matrix $H^{\rm PM}_1(t)$ for all $t$, 
evolves faster than $u_2(x,t) \in {\cal M}$,
while $\hat u_1(x,t)$ and $u_2(x,t)$ have the common trajectory 
on ${\cal M}$ by Theorem \ref{thm1} 
(See Figure \ref{fig3}). 
The corollary suggests that by measuring the diagonal elements of 
$H_1^{\rm PM}(t)$ we can estimate how far
$u_1(x,t)$ is from ${\cal M}$ in terms of the divergence.
Note that the difference of velocities vanishes when $m \rightarrow 1$.
Hence, this is the specific property of the slow diffusions governed by 
the PME.

Let $\eta^{\rm FP}(\tau)$ and $H^{\rm FP}(\tau)$ be, respectively, 
the first and the second moments of $p(x,\tau)$, i.e.,
\[
	\eta^{\rm FP}= \mathbf{E}_p[x], 
	\quad H^{\rm FP}=\mathbf{E}_p[xx^T].
\]
From the behavior of the moments of the PME and 
the above relations of moments, we have 
\begin{eqnarray*}
	\eta^{\rm FP}(\tau)
	&=& e^{-\beta \tau}\eta^{\rm FP}(0), \\
	H^{\rm FP}(\tau)
	&=& e^{-2\beta \tau}H^{\rm FP}(0)
		+ e^{-2\beta \tau} \sigma^{\rm FP}_p(e^\tau-1)D,
\end{eqnarray*}
where the scaling $\tau=\ln(t+1)$ is assumed 
and $\sigma^{\rm FP}_p(t)$ is defined by
\begin{eqnarray*}
	\sigma^{\rm FP}_p(t)
		&:=&2\int_0^{\ln(1+t)} d\tau' e^{\tau'+\mu(1-m)\tau'} 
			\int p(x,\tau')^m dx \\
		&=& \det(R) \sigma^{\rm PM}_u(t).
\end{eqnarray*}
for a solution $u$ of the PME and the corresponding solution $p$ of the NFPE.
Note that differentiating the above by $t$, we have the relation:
\begin{equation}
	(1+t)^{\mu(1-m)} \int p(z,\tau)^m dz
	= \det (R) \int u(x,t)^m dx.
\label{m_trans}
\end{equation}

For the limiting case $m \rightarrow 1$ (and accordingly 
$\beta \rightarrow 1/2$),
we see that the above expressions recover the well-known 
linear Fokker-Plank case with a drift vector $x/2$:
\begin{eqnarray*}
		\eta^{\rm FP}(\tau)&=&e^{- \tau/2}\eta^{\rm FP}(0), \\
		H^{\rm FP}(\tau)&=&e^{- \tau}H^{\rm FP}(0)
		+ 2(1-e^{-\tau})D.
\end{eqnarray*}
Since we know that $p(x,\tau)$ converges to $p_\infty(x) \in {\cal M}$ 
in (\ref{eq_density}) and it holds that
\begin{equation}
	\lim_{\tau \rightarrow \infty} H^{\rm FP}(\tau)
	=\sqrt{\det D} \left(\lim_{t \rightarrow \infty} 
	(t+1)^{-2\beta}\sigma^{\rm PM}_u(t) \right)D
\label{limitingcase}
\end{equation}
because $\det R=\sqrt{\det D}$, we conclude that 
the left-hand side of (\ref{limitingcase}) exists and 
$\sigma^{\rm PM}_u(t)=O(t^{2\beta})$ as $t \rightarrow \infty$ 
(Cf. Remark \ref{rem_thm}).
Summing up the above with Proposition \ref{prop_geod}, 
we obtain the following geometric property of the NFPE:
\begin{corollary}
Consider solutions of the NFPE 
with the common initial first and second moments.
Then their m-projections to ${\cal M}$ evolve along 
with the common m-geodesic curve approaching from the density determined 
by the initial moments to the equilibrium $p_\infty(x)$.
\end{corollary}

\subsection{Convergence rate of the solution of the PME to ${\cal M}$}
The following proposition is obtained from ii) of Proposition \ref{prop_proj} 
and a result \cite{CT,Toscani} claiming that a solution of the NFPE 
decays exponentially with respect to the divergence, i.e.,
\begin{equation}
	{\cal D}[p(x,\tau) \| p_\infty(x)]
	\le  {\cal D}[p(x,0) \| p_\infty(x)]e^{-2 \beta \tau}.
\label{expdecay}
\end{equation}

\begin{proposition}
\label{rate2M}
Let $u(x,t)$ be a solution of the PME and $\hat u(x,t)$ be the m-projection 
of $u(x,t)$ to the $q$-Gaussian family ${\cal M}$ at each $t$.
Then $u(x,t)$ asymptotically approaches to ${\cal M}$ with
\[
	{\cal D}[u(x,t) \| \hat u(x,t)] \le \frac{C_0}{1+t},
\]
where $C_0$ is a constant depending on the initial function $u(x,0)$.
\end{proposition}
\begin{remark}
Combining this result and 
the Csiszar-Kullback inequality \cite{CT}, 
we can also conclude the $L^1$ convergence 
of $u(x,t)$ to ${\cal M}$ with the rate $1/\sqrt{1+t}$.
This implies that the convergence to ${\cal M}$ is faster than $1/t^\beta$, 
which is the convergence rate to the self-similar solution of the PME in the 
case $1<m \le 2$ \cite{Vaz,Toscani}.
\end{remark}

\section{Conclusions}

We have discussed the Legendre duality of generalized entropies and 
its applications focusing on the duality relation (\ref{drel}) 
of the statistical manifold.
Within this framework, we classified extensions into two major directions 
using the representation functions.
In terms of corresponding geometric structure, we can say that 
the one is characterized 
by nonflatness with constant curvature, while the other is by dual flatness.

The important point would be how such a geometric setup is useful 
for our understandings of various physical phenomena not obeying 
the standard statistical theory.
For this purpose we have partially presented the recent results on 
the solutions of the PME and the NFPE.
However, the whole picture of the proposed generalization in Section \ref{GCE} 
is still largely formal and it needs more developments on the basis of 
physical background.

%
%
%
%
%
%

%
%

\begin{appendix}
\noindent {\bf\Large Appendix}
\section{Statistical manifolds and affine differential geometry}
In this appendix we briefly summarize several concepts and properties 
of statistical manifolds and affine hypersurfaces, which are necessary 
in this paper.
See for details \cite{Kurose1,Kurose2,NS}, respectively.
\subsection{Statistical manifold}
For a torsion-free affine connection $\nabla$ and a pseudo Riemannian 
metric $g$ on a manifold $\mathcal{M}$, the triple $(\mathcal{M},g,\nabla)$ 
is called a {\em statistical manifold} if 
it admits another torsion-free connection $\nabla^*$ satisfying
\begin{equation}
 Xg(Y,Z)=g(\nabla_X Y, Z)+g(Y, \nabla^*_X Z)
\label{dualrel}
\end{equation}
for arbitrary $X,Y$ and $Z$ in ${\mathcal X}(\mathcal{M})$, where ${\mathcal X}(\mathcal{M})$ is the set of all tangent vector fields on $\mathcal{M}$.
We call $\nabla$ and $\nabla^*$ {\em duals} of each other with respect to $g$, 
and $({\cal M},g,\nabla^*)$ is said dual statistical manifold 
of $({\cal M},g,\nabla)$.
The triple of a Riemannian metric and a pair of dual connections 
$(g,\nabla,\nabla^*)$ satisfying (\ref{dualrel}) is called 
{\em dualistic structure} on $\mathcal{M}$, 
which plays a fundamental role in the study of manifolds of 
probability distributions.

A statistical manifold $(\mathcal{M},g,\nabla)$ is said to have {\em constant
curvature} $\kappa$ if the curvature tensor $R$ of $\nabla$ satisfies
\begin{equation}
 R(X,Y)Z=\kappa \{g(Y,Z)X-g(X,Z)Y\}.
\label{smcc}
\end{equation}
When the constant $\kappa$ is zero, the statistical manifold is said to be
{\em flat}, or {\em dually flat}, because the curvature tensor $R^*$ 
of $\nabla^*$ is known to vanish automatically.


\subsection{Affine hypersurface theory}
Let $\mathcal{M}$ be an $n$-dimensional manifold and consider 
an {\em affine immersion} $(f,\xi)$, which is the pair of 
an immersion $f$ of $\mathcal{M}$ into $\mathbf{R}^{n+1}$ and a
transversal vector field $\xi$ to $f({\cal M})$.
We denote by $D_X f_*(Y)$ the covariant derivative along $f$ induced by 
the standard flat connection $D$ of $\mathbf{R}^{n+1}$.
By a given affine immersion $(f,\xi)$ of $\mathcal{M}$, 
the Gauss and Weingarten formulas are respectively obtained as follows:
\begin{eqnarray}
 D_Xf_*(Y) &=& f_*(\nabla_X Y) + h(X,Y)\xi, \label{Gauss} \\
 D_X \xi &=&- f_*(SX) + \tau(X) \xi. \label{Weingarten}
\end{eqnarray}
Here, $\nabla, h, S$ and $\tau$ determined from the above formulas 
are called, respectively, the {\em induced
connection, affine fundamental form, affine shape operator} and
{\em transversal connection form} \cite{NS}.
By regarding $h$ as a (pseudo-) Riemannian metric of $\mathcal{M}$, 
we say that the affine immersion realizes
$(\mathcal{M},h,\nabla)$ in $\mathbf{R}^{n+1}$.

An affine immersion is said {\em nondegenerate} and {\em equiaffine} 
if $h$ is nondegenerate and $\tau=0$, respectively.
Further, let $o$ be the origin of ${\bf R}^{n+1}$.
Then we say that the affine immersion is 
{\em centro-affine with a scaling constant $\rho$} if 
$\xi$ at $f(x)$ is equal to $-\rho$ times of the vector 
$\overline{of(x)}^\triangleright$
for a constant $\rho$ and $x \in {\cal M}$.


The following facts hold \cite{NS}, which are convenient to know 
the property of the realized manifold by an affine immersion:
\begin{description}
\item[F1)] An equiaffine and nondegenerate affine immersion realizes 
a statistical manifold,
\item[F2)] A centro-affine immersion with a scaling constant $\rho$ 
is equiaffine with $S=\rho I$.
The realized manifold is projectively flat, 
\item[F3)] An equiaffine and nondegenerate affine immersion 
with $S=\rho I$ realizes a statistical manifold 
with constant curvature $\rho$.
\end{description}

Let $(\mathcal{M}, g, \nabla)$ be a statistical manifold with constant
curvature, which is realized by a centro-affine immersion with constant 
scaling. 
It is known that $(\mathcal{M},g,\nabla)$ has the following properties 
\cite{Kurose2,NS}:
\begin{description}
 \item[P1)] Modified Pythagorean relation for the geometric divergence 
		(contrast function) holds on $\mathcal{M}$,
 \item[P2)] Let $\gamma$ be a $\nabla$-geodesic curve joining two points 
		$x$ and $y$ on $\mathcal{M}$.
		Then, $f(\gamma)$ is the intersection of 
		$f(\mathcal{M})$ and the two-dimensional plane containing
		$\overline{of(x)}^\triangleright$ and 
		$\overline{of(y)}^\triangleright$ in ${\bf R}^{n+1}$.
\end{description}
The corresponding results also hold for the dual statistical manifold 
$(\mathcal{M}, g, \nabla^*)$ with constant curvature.

\section{Proof of Proposition \ref{prop_com}}

First, recall that $D^{(\alpha)}$ is the restriction of ${\cal D}$ 
in (\ref{alphadiv}) defined on ${\bf R}^{n+1}_+ \times {\bf R}^{n+1}_+$.
Hence, in the ambient space $({\bf R}^{n+1}_+, \tilde{g}, D)$, 
the gradient vector $\tilde{X}$ of 
${\cal D}(\bm{\theta},\bm{\theta}(\bm{r}))$ at 
$\bm{\theta} \in {\bf R}^{n+1}_+$
is represented by
\[
\tilde{X}=\sum_{i=1}^{n+1} \tilde{X}^i \frac{\partial}{\partial \theta^i}, 
\]
where
\begin{eqnarray*}
\tilde{X}^i &=& -\sum_{j=1}^{n+1} \tilde{g}^{ij} 
\frac{\partial}{\partial \theta^j} {\cal D}(\bm{\theta},\bm{\theta}(\bm{r})) \\
&=&\sum_{j=1}^{n+1} \tilde{g}^{ij}(\eta_j(\bm{r})-\eta_j), 
\; i=1,\cdots,n+1,
\end{eqnarray*}
and $\tilde{g}^{ij}$ is the component of the inverse of the Riemannian 
metric $\tilde{g}=(\partial^2 \psi/\partial \theta^i \partial
\theta^j)$ given in (\ref{Hess}).
Hence, the gradient vector $\tilde X$ is 
represented in the $\eta$-coordinates by
\[
\tilde{X}=\sum_{i=1}^{n+1} \tilde{X}_i \frac{\partial}{\partial \eta_i}, 
\quad \tilde X_i=\eta_i(\bm{r})-\eta_i
\]
and its integral curve is
\[
	\eta_i(\bm{p}(t))=e^{-t}(\eta_i(\bm{p}_0)-\eta_i(\bm{r}))+\eta_i(\bm{r}).
\]

Next, consider the vector field defined by
\[
 N=\sum_{i=1}^{n+1} p_i \frac{\partial}{\partial p_i} 
= \sum_{i=1}^{n+1} \frac{1-\alpha}{2}\theta^i \frac{\partial}{\partial
\theta^i}
= \sum_{i=1}^{n+1} \frac{1+\alpha}{2}\eta_i \frac{\partial}{\partial
\eta_i},
\]
then $N$ is orthogonal to $T_{\bms{p}} \mathcal{S}^n$ at
each $\bm{p} \in \mathcal{S}^n$ with respect to $\tilde{g}$.
It is verified that the gradient vector $X$ 
is represented as the orthogonal projection of 
$\tilde{X}$ onto $T_{\bms{p}} \mathcal{S}^n$ along $N$.

Combining the above two facts, we see that 
the integral curve (gradient flow) of $X$ is restricted to the
two-dimensional plane $\mathcal{P}^*$ containing the three points 
$\bm{\eta}(\bm{p}_0)$, $\bm{\eta}(\bm{r})$ and 
the origin $o$ in $({\bf R}^{n+1})^*$ 
with the $\eta$-coordinate system.
Thus, the second statement follows.
Note that ${\cal S}^n$ is represented as the level surface 
$\varphi(\bm{\eta})=2/(1-\alpha)$ in $({\bf R}^{n+1})^*$, then 
we conclude that the gradient flow is actually represented by 
the intersection of the level surface and $\mathcal{P}^*$.
This proves the first statement owing to the property (P2) in Appendix A 
for the dual statistical manifold $({\cal S}^n,g,\nabla^{(-\alpha)})$. 
\qed
\end{appendix}
\end{document}